\documentclass[conference]{IEEEtran}
\IEEEoverridecommandlockouts
\usepackage{cite}
\usepackage{amsmath,amssymb,amsfonts}
\usepackage{algorithmic}
\usepackage{graphicx}
\usepackage{textcomp}
\usepackage{subcaption}
\usepackage{xcolor}
\def\BibTeX{{\rm B\kern-.05em{\sc i\kern-.025em b}\kern-.08em
    T\kern-.1667em\lower.7ex\hbox{E}\kern-.125emX}}
\begin{document}

\title{A General Close-loop Predictive Coding Framework for Auditory Working Memory\\
}

% email address or ORCID}}
\author{Zhongju Yuan\textsuperscript{1}, Geraint Wiggins\textsuperscript{2, 3 *}, Dick Botteldooren\textsuperscript{1*}\thanks{*Corresponding author: Geraint Wiggins, Dick Botteldooren.}\\
  \textsuperscript{1}WAVES Research Group, Ghent University, Gent, Belgium\\
  \textsuperscript{2}AI Lab, Vrije Universiteit Brussel, Belgium\\
  \textsuperscript{3}EECS, Queen Mary University of London, UK\\
  \texttt{zhongju.yuan@ugent.be},
\texttt{geraint.wiggins@vub.be},
\texttt{dick.botteldooren@ugent.be}\\
}

\maketitle

\begin{abstract}
Auditory working memory is essential for various daily activities, such as language acquisition, conversation. It involves the temporary storage and manipulation of information that is no longer present in the environment. While extensively studied in neuroscience and cognitive science, research on its modeling within neural networks remains limited. To address this gap, we propose a general framework based on a close-loop predictive coding paradigm to perform short auditory signal memory tasks. The framework is evaluated on two widely used benchmark datasets for environmental sound and speech, demonstrating high semantic similarity across both datasets.
\end{abstract}


\section{Introduction}

Auditory working memory plays a critical role in various daily activities, including language learning, conversation, and note-taking. It refers to the ability to temporarily store and manipulate information that is no longer present in the environment~\cite{kaiser2015dynamics}. Despite its significance, the neural mechanisms underpinning auditory working memory—particularly the processes enabling the active maintenance of sounds over short periods—remain an area of active investigation~\cite{kumar2016brain}.

Numerous artificial intelligence models have been developed to address memory-related tasks~\cite{muezzinoglu2003new, kohonen2012associative}. However, many of these models are primarily designed for pattern storage and retrieval~\cite{lowe1999storage, krotov2016dense}, rendering them less effective for tasks requiring the sequential encoding and retrieval of information. In contrast, auditory working memory demands the capability to encode and reconstruct entire auditory sequences based on a given cue from the initial segment.

Predictive coding offers a biologically inspired approach to sequential memory encoding~\cite{quak2015multisensory, alexander2018frontal}. Its top-down mechanism closely parallels backpropagation in artificial intelligence. In this work, we employ a two-layer system with distinct sets of weights to predict the subsequent hidden states and observations. This approach allows the retrieval of auditory sequences by leveraging the parameters encoded during the initial cue.

While previous studies have explored sequential memory using predictive coding~\cite{tang2024sequential}, most research has focused on visual memory~\cite{millidge2022universal, lu2024learning}. This emphasis is partly due to the relative simplicity of evaluating visual tasks, as visual features are more easily interpretable. In contrast, auditory sequences present complex temporal and spectral features that cannot be readily discerned by the human eye.

Our preliminary experiments indicate that due to the intricate nature of auditory segment information, models relying solely on an initial cue often fail to recall sequences accurately. A close-loop approach, which incorporates feedback, significantly enhances retrieval accuracy. Feedback loops are critical for maintaining internally generated cognitive representations in biological systems~\cite{lim2013balanced, voitov2022cortical}. Motivated by this, we propose a general framework for auditory memory tasks based on close-loop predictive coding.

Given the challenge of visually interpreting audio recall results, we employ well-established audio recognition models to generate textual captions for the recalled sequences. This method simulates verbal recall, a commonly used approach in neuroscience studies~\cite{lenzenweger2000auditory}.

In this work, we introduce a close-loop predictive coding framework for auditory working memory. We evaluate memory and recall performance using two audio datasets: one comprising environmental sounds and another focused on speech. The recall accuracy is assessed by comparing the textual captions generated for the original and recalled audio using models for environmental sound captioning and speech recognition.

Our primary contributions are summarized as follows:
\begin{itemize} 
\item We propose a general close-loop predictive coding method for auditory working memory, incorporating a biologically inspired feedback mechanism to enhance sequence recall accuracy.
\item We design a robust framework to evaluate recall results by textual similarity (via audio captioning and speech recognition).
\item We validate the proposed method on two diverse datasets—environmental sounds and speech—demonstrating its effectiveness in achieving high memory performance across different auditory contexts.
\end{itemize}

\section{Related Work}

Auditory working memory, a critical component of the human brain's cognitive model, has been extensively studied across diverse tasks, including music~\cite{schulze2012working,hansen2013working,fernandez2022associations} and speech processing~\cite{rudner2012working, gordon2016effects}. These investigations predominantly rely on neuroimaging techniques to decode auditory working memory functions~\cite{kaiser2015dynamics, kumar2016brain, wilsch2016works,czoschke2021decoding}.

In~\cite{yu2021causal}, the causal relationship between the auditory cortex and auditory working memory was investigated using functional magnetic resonance imaging (fMRI). Similarly, the work in~\cite{albouy2022supramodality} utilized MEG/EEG techniques to demonstrate that theta-rhythmic visual stimulation during auditory memory tasks causally enhances performance. Key drivers of this effect were identified as the stimulus's rotating properties and flickering frequency. Additionally,~\cite{fernandez2022associations} revealed that individuals with high working memory performance engaged a broader brain network, including regions associated with visual processing—such as the inferior temporal, temporal-fusiform, and postcentral gyri—enabling successful auditory memory recognition.

Predictive coding has been identified as a critical mechanism in auditory working memory tasks~\cite{fernandez2022associations}. The study in~\cite{fernandez2022associations} highlighted that high-beta/low-gamma EEG activity reflects top-down predictive coding during spatial working memory tasks, illustrating how the brain minimizes prediction errors by reconciling predictions with sensory inputs. Neuroimaging findings in~\cite{alexander2018frontal} further support this framework, demonstrating that prefrontal cortex activity aligns with hierarchical prediction and error-correction processes. Specifically, regions such as the dorsolateral and medial prefrontal cortex compute and maintain progressively abstract error signals to guide adaptive behavior. The hierarchical error representation (HER) model effectively simulates these findings, reinforcing predictive coding as a unifying principle for neural and cognitive processes.

Close-loop feedback plays a critical role in working memory. In~\cite{lim2014balanced}, the authors demonstrated that mechanisms such as negative-derivative feedback are essential for maintaining persistent neural activity in working memory tasks. These mechanisms mitigate memory decay, balance excitation and inhibition, and enable robust encoding of both spatial and amplitude information. Their findings align with experimental observations in the prefrontal cortex and illustrate how such feedback enhances the system's resilience to perturbations, underscoring their biological relevance. Similarly,~\cite{voitov2022cortical} emphasizes the significance of reciprocal cortical feedback loops in sustaining working memory. Their study highlights how distributed neocortical regions leverage high-dimensional representations, facilitated by feedback interactions, to ensure the persistence and accuracy of sensory information, thereby playing a pivotal role in encoding and retrieving working memory-specific behavioral responses.

Several studies in artificial intelligence have explored neural network-based approaches to sequential memory modeling~\cite{spratling2016predictive}. For instance, \cite{tang2024sequential} introduces a predictive coding-based model specifically designed for visual sequential memory. Similarly, \cite{chaudhry2024long} presents a Hopfield network model enhanced with a nonlinear interaction term, significantly increasing its sequence capacity. Moreover, \cite{spens2024generative} proposes a generative framework to tackle the reconstruction task, focusing on episodic memory development. However, these methods are predominantly tailored for visual memory applications.

\section{Methods}
\begin{figure*}[t]
    \centering
    \includegraphics[width=\linewidth]{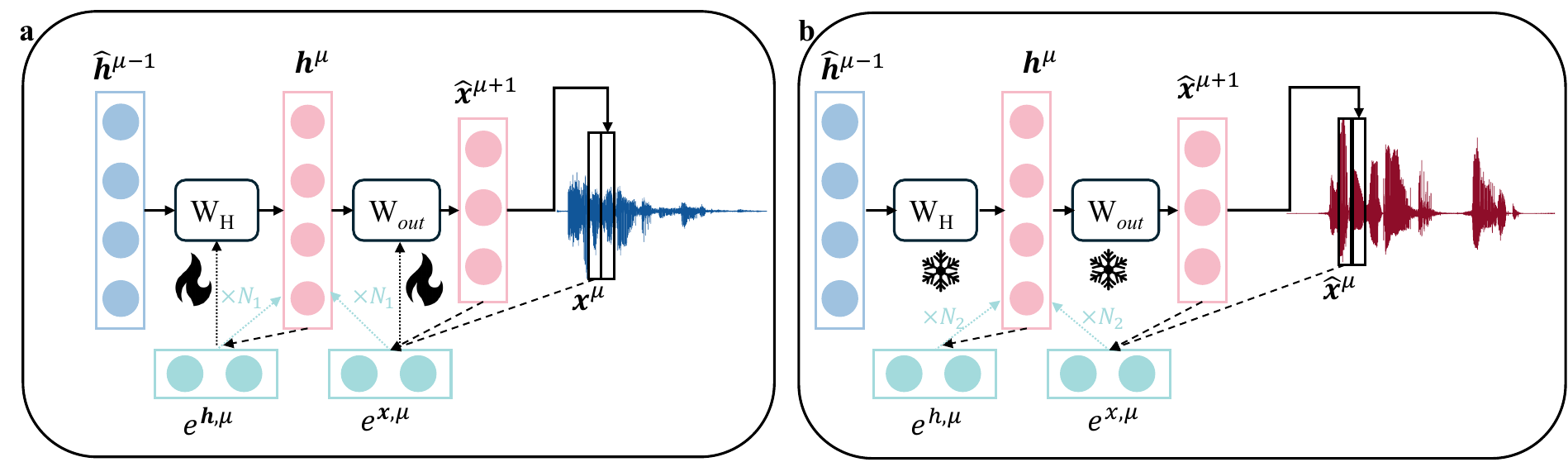}
    \caption{\textbf{Memory Procedure for an Audio Sequence.} (a) shows the \textbf{write} process, where weight matrices are trained in a supervised manner to minimize error terms. (b) illustrates the \textbf{read} process, during which the trained weights remain fixed. The model uses the recalled last segment to update the hidden state, enhancing the quality of the subsequent segment retrieval.}
    \label{fig:overview}
\end{figure*}

\begin{figure*}[t]
    \centering
    \begin{subfigure}[b]{0.48\textwidth}
        \centering
        \includegraphics[width=\textwidth]{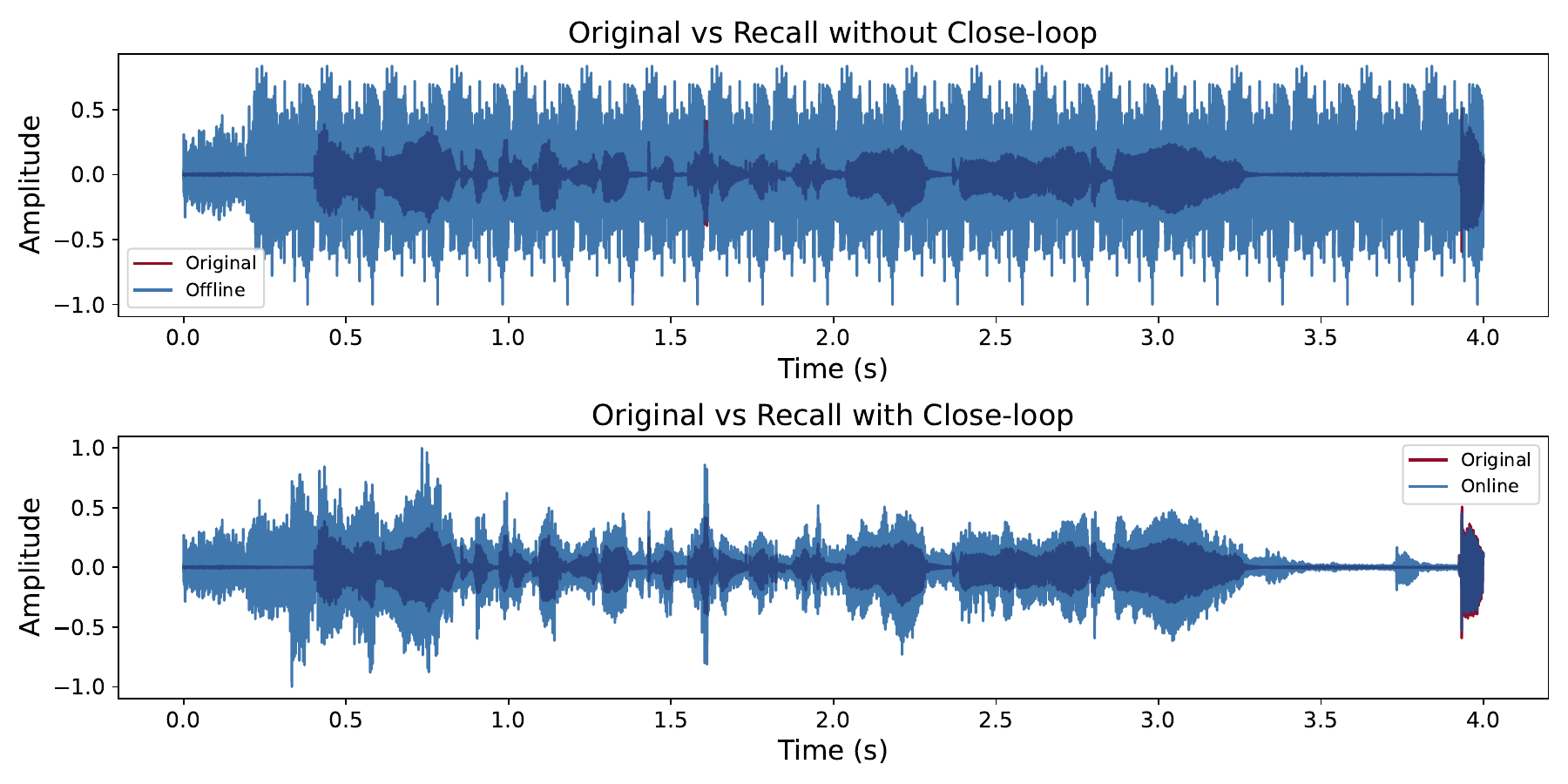}
        \caption{Waveform comparison for speech example 1.}
        \label{fig:speech1_close_loop}
        % : `I had wanted to get some picture books for Yulka and Antonia.`
    \end{subfigure}
    \hfill
    \begin{subfigure}[b]{0.48\textwidth}
        \centering
        \includegraphics[width=\textwidth]{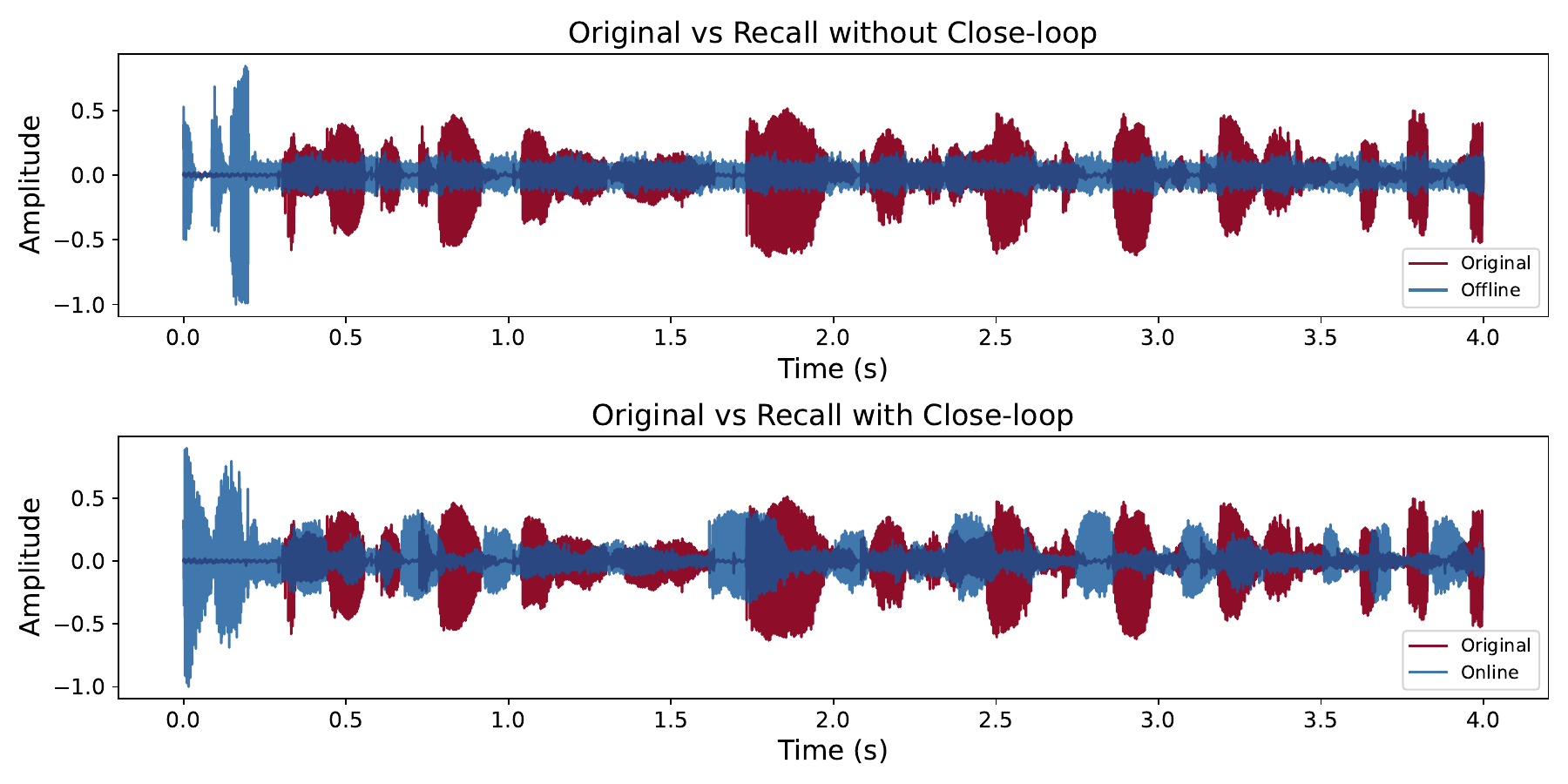}
        \caption{Waveform comparison for speech example 2.}
        % : `I'm not a talker, you know, and, as the laws of gravitation forbid my soaring aloft anywhere.`
        \label{fig:speech2_close_loop}
    \end{subfigure}
    \caption{\textbf{Waveform comparisons of original and recalled audio for two speech examples.} The upper panels show the recalled waveforms without using the close-loop approach, while the lower panels show the recalled waveforms using the close-loop approach. The close-loop method significantly improves waveform accuracy and alignment, resulting in consistent recognition performance.}
    \label{fig:close_loop_figures}
\end{figure*}

\begin{figure*}[t]
    \centering
    \begin{subfigure}[b]{0.48\textwidth}
        \centering
        \includegraphics[width=\textwidth, trim={0 0 120 85}, clip]{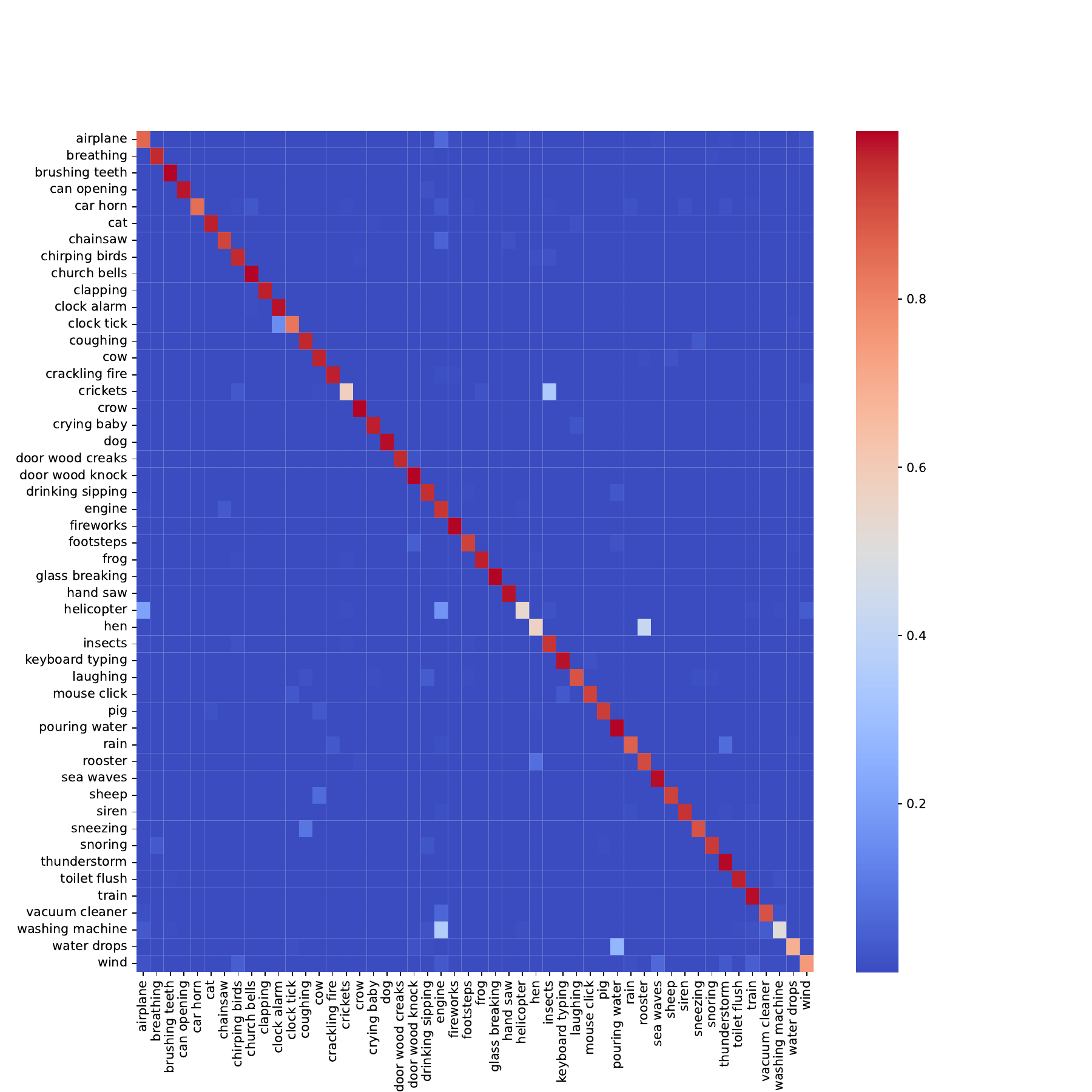}
        \caption{Classification probability distribution of ground truth dataset.}
        \label{fig:ori_close_loop}
    \end{subfigure}
    \hfill
    \begin{subfigure}[b]{0.48\textwidth}
        \centering
        \includegraphics[width=\textwidth, trim={0 0 120 85}, clip]{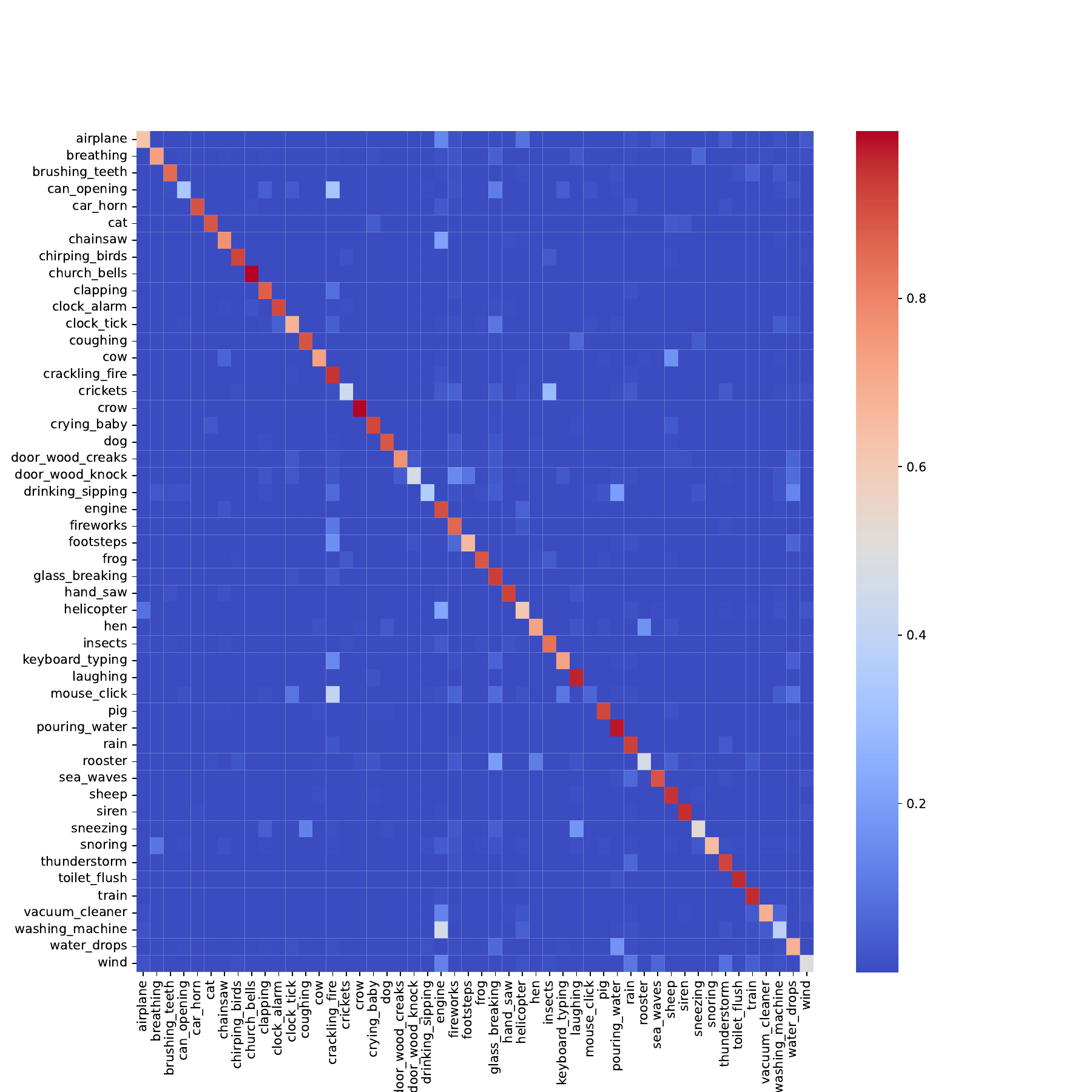}
        \caption{Classification probability distribution of recalled dataset.}
        \label{fig:recall_close_loop}
    \end{subfigure}
    \caption{\textbf{Classification probability distribution heatmaps for the original and recalled audio in the \texttt{ESC-50} dataset.} (a) A classification probability heatmap for the original audio demonstrates high accuracy, with diagonal elements nearing 1. However, a few semantically ambiguous classes, such as `helicopter' and `airplane' or `crickets' and `insects', show slight deviations. (b) Classification probability heatmap for the recalled audio, where most diagonal elements retain probabilities above 0.7.}
    \label{fig:close_loop_figures_esc}
\end{figure*}

\begin{figure*}[t]
    \centering
    \includegraphics[width=\linewidth]{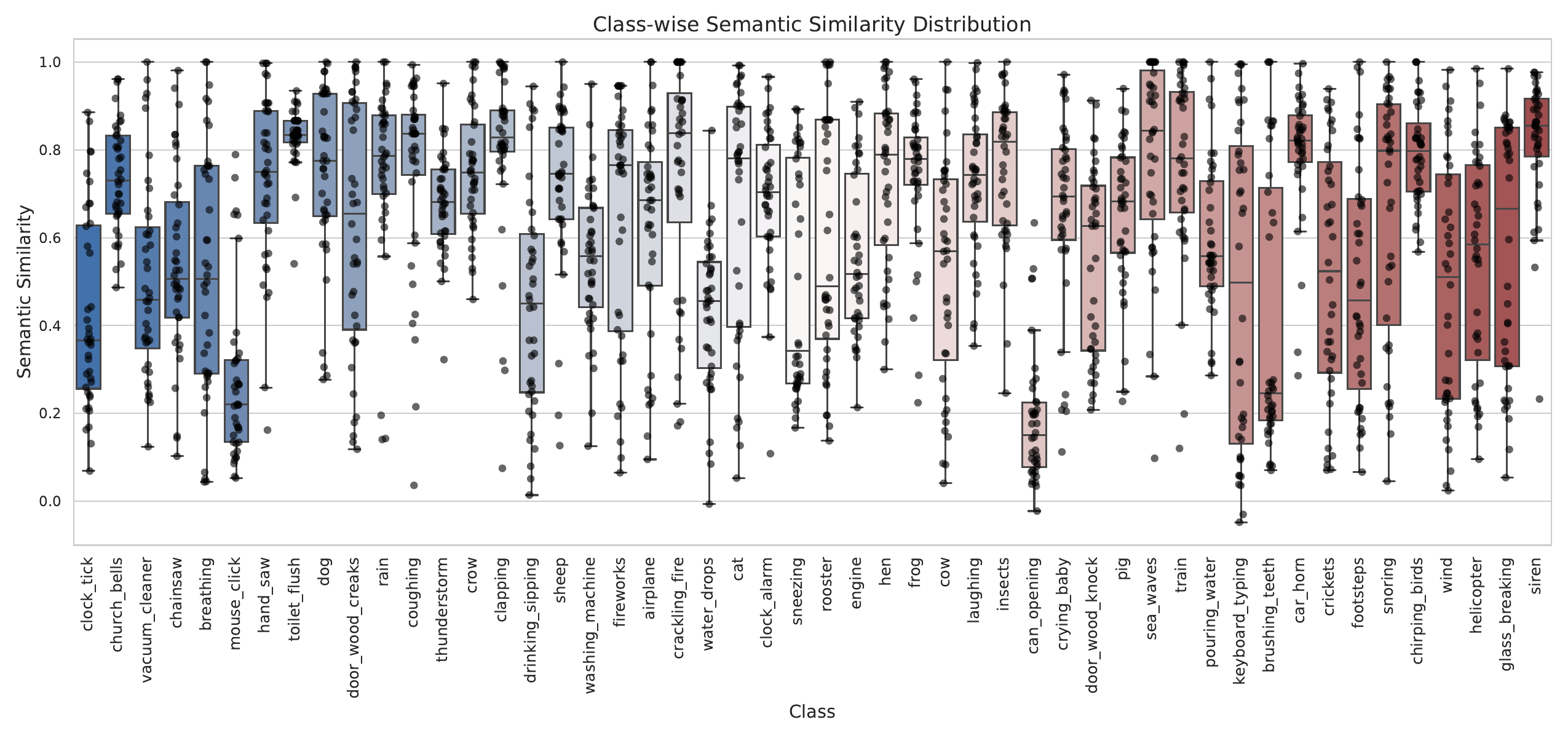}
    \caption{\textbf{Performance analysis of the environmental sound memory model.} Boxplot of semantic similarities (SS) between original and recalled audio pairs in the \texttt{ESC-50} dataset. The x-axis represents the 50 class names, while the y-axis shows the semantic similarity scores. Black dots indicate the distribution of similarity scores for each class.}
    \label{fig:ESC_results}
\end{figure*}

\begin{figure*}[t]
    \centering
    \includegraphics[width=\linewidth]{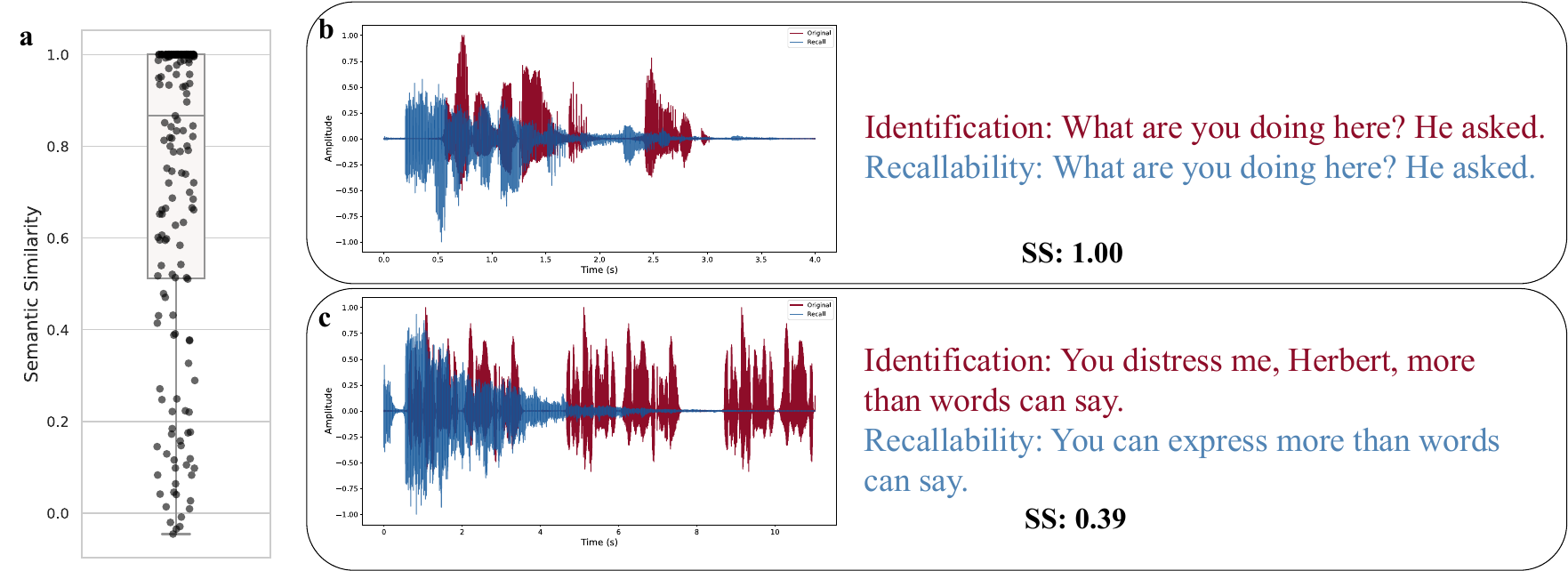}
    \caption{\textbf{Performance Analysis of the Speech Memory Model.} The identification represents the Whisper recognition result for the ground truth audio, while recallability denotes the result obtained from the same model for the recalled audio. Panel (a) displays the semantic similarity (SS) as a box plot, highlighting variations in performance. Panel (b) illustrates cases with high SS, indicating successful retention of both semantic and acoustic features. In contrast, Panel (c) presents a negative sample with lower SS, demonstrating a failure to retain meaning despite preserving some degree of acoustic fidelity.}
    \label{fig:speech_results}
\end{figure*}

The task involves auditory sequence memory. Consider an auditory sequence divided into segments of length 10 ms. Let each segment be indexed by $i$, where $i$ ranges from $1$ to $L/l$, with $l$ representing the segment length. The index of the current segment is denoted as $\mu$.

The model consists of a single hidden layer with two general weight matrices: $\mathbf{W}_{H}$ and $\mathbf{W}_{out}$. The hidden state at segment $\mu$ is denoted as $\mathbf{h}^{\mu}$, and the corresponding output is $\hat{\mathbf{x}}^{\mu}$. The matrix $\mathbf{W}_{H}$ is utilized to predict the subsequent hidden state $\mathbf{h}^{\mu+1}$, while $\mathbf{W}_{out}$ is employed to predict the next output $\hat{\mathbf{x}}^{\mu+1}$. The loss function is defined as the $\ell_2$-norm of the prediction errors for both weight matrices, as expressed in Eq.~(\ref{eq:loss_function_write}). The procedure is divided into two phases: \textbf{write} and \textbf{read}.
\begin{equation}
\begin{aligned}
    \mathcal{F}_{\mu} &= \|e^{\mathbf{x}, \mu}\|^2 + \|e^{\mathbf{h}, \mu}\|^2 \\
    &= \|\mathbf{x}^{\mu} - \mathbf{W}_{out}f(\mathbf{h}^{\mu})\|_2^2 + \|\mathbf{h}^{\mu} - h(\mathbf{W}_{H}\hat{\mathbf{h}}^{\mu-1})\|_2^2,
\end{aligned}
\label{eq:loss_function_write}
\end{equation}
where $f(\cdot)$ and $h(\cdot)$ are non-linear functions.

During the \textbf{write} procedure (shown in Fig.~\ref{fig:overview}(a)), the current audio segment $\mathbf{x}^{\mu}$ is written into the weight matrices $\mathbf{W}_{H}$ and $\mathbf{W}_{out}$. The hidden state at the current segment, $\Tilde{\mathbf{h}}^{\mu}$, is iteratively updated during this process until the weight matrices converge. The update of $\Tilde{\mathbf{h}}^{\mu}$ in the \textbf{write} phase is performed using the partial differential gradient, as shown in Eq.~(\ref{eq:update_read})\cite{tang2024sequential}.
\begin{equation}
    \Tilde{\mathbf{h}}^{\mu} \propto -\frac{\partial \mathcal{F}_{\mu}}{\partial \mathbf{z}^{\mu}}
    \propto - e^{\mathbf{h}, \mu} + f'(\mathbf{h}^{\mu})\odot \mathbf{W}_{out}^\top e^{\mathbf{x}, \mu}
\label{eq:update_read}
\end{equation}

The training of $\mathbf{W}_H$ and $\mathbf{W}_{out}$ is governed by the backpropagation rule, as expressed in Eq.(\ref{eq:train_WF}) and Eq.(\ref{eq:train_WH}).
\begin{equation}
\Delta \mathbf{W}_{out} \propto -\frac{\partial \mathcal{F}_\mu (\mathbf{h}^\mu, \mathbf{W}_H, \mathbf{W}_{out})}{\partial \mathbf{W}_{out}} = e^{x, \mu} f(\mathbf{h}^\mu)^\top.
\label{eq:train_WF}
\end{equation}
\begin{equation}
\begin{aligned}
    \Delta \mathbf{W}_H \propto -\frac{\partial \mathcal{F}_\mu (\mathbf{h}^\mu, \mathbf{W}_H, \mathbf{W}_{out})}{\partial \mathbf{W}_H} 
    = e^{\mathbf{h}, {\mu}} h'(\mathbf{W}_{H}\hat{\mathbf{h}}^{\mu-1})\mathbf{h}^{\mu-1}
\end{aligned}
\label{eq:train_WH}
\end{equation}

After completing a predefined number of epochs, the auditory memory is encoded in $\mathbf{W}_{H}$ and $\mathbf{W}_{out}$. During the \textbf{read} procedure, these matrices, $\mathbf{W}_{H}$ and $\mathbf{W}_{out}$, are fixed and no longer updated, as depicted in Fig.\ref{fig:overview}(b). The retrieval of auditory segments is guided by the loss function defined in Eq.\ref{eq:loss_function_read}:
\begin{equation}
\mathcal{F}_\mu(\mathbf{h}^\mu, \hat{\mathbf{x}}^\mu) = \|\mathbf{h}^\mu - \mathbf{W}_H f(\hat{\mathbf{h}}^{\mu-1})\|_2^2 + \|\hat{\mathbf{x}}^\mu - \mathbf{W}_{out} f(\mathbf{h}^\mu)\|_2^2
\label{eq:loss_function_read}
\end{equation}

The iterative process for extracting the segment at $\mu$, denoted as $\hat{\mathbf{x}}^\mu$, is expressed in Eq.~(\ref{eq:read_segment}):
\begin{equation}
\hat{\mathbf{x}}^\mu \propto -\frac{\partial \mathcal{F}_\mu(\mathbf{h}^\mu, \hat{\mathbf{x}}^\mu)}{\partial \hat{\mathbf{x}}^\mu} \propto -\boldsymbol{e}^{\mathbf{x}, \mu}
\label{eq:read_segment}
\end{equation}

The hidden state is updated iteratively using the error terms defined in Eq.~(\ref{eq:loss_function_write}), following the update procedure outlined in Eq.~(\ref{eq:update_read}).

Given the complexity and length of the auditory signal, a close-loop feedback mechanism is incorporated during the \textbf{read} procedure. This feedback introduces differences in how the loss terms are calculated between the \textbf{write} and \textbf{read} procedures.

During the \textbf{write} phase, $e^{\mathbf{x}, \mu}$ is computed as the Euclidean distance between the segment $\mathbf{x}^{\mu}$ to be memorized and the corresponding segment stored in the weight matrices after the iteration, $\hat{\mathbf{x}}^{\mu}$. In contrast, during the \textbf{read} phase, $e^{\mathbf{x}, \mu}$ is determined as the Euclidean distance between the segment stored in the weight matrices after the iteration, $\hat{\mathbf{x}}^{\mu}$, and the most recently recalled audio segment.

\section{Experiments}
In this section, we evaluate the performance of the proposed framework on two public audio datasets, \texttt{ESC-50} and \texttt{LibriSpeech}. Specifically, we apply our method to each audio sample in these datasets to determine whether the framework achieves its objective of memorizing the audio sequence. To provide a thorough comparison, we measure the memorizing performance by two state-of-the-art models, employing semantic similarity (SS) as an evaluation metric for both datasets.

\subsection{Dataset}
This study utilizes two widely recognized audio datasets: \texttt{ESC-50}~\cite{piczak2015esc} for environmental sound classification and \texttt{LibriSpeech}~\cite{panayotov2015librispeech} for speech-related tasks.

The \texttt{ESC-50} dataset comprises 2,000 audio samples spanning 50 classes, grouped into five categories, including animal sounds, natural ambient sounds, and urban noises. Each clip is 5 seconds long and sampled at 44.1 kHz.

The \texttt{LibriSpeech} dataset, derived from audiobooks, contains approximately 1,000 hours of English speech with high-quality transcriptions. It is divided into training, development, and test sets, with subsets categorized by difficulty. In this study, we extract 180 clean samples in \texttt{LibriSpeech} to evaluate the recall accuracy. 

\subsection{Experimental settings}
The audio files are processed at a resampling rate of 16 kHz for \texttt{ESC-50}, and 44.1 kHz for \texttt{LibriSpeech}. Instead of processing each individual sampling point, the model processes 200 ms segments sequentially to manage computational complexity and reduce the memory requirements. For each audio sample, the model retains information for the first 20 steps (equivalent to 4 seconds) and performs the writing operation for 100 epochs. The loss function used for both observation prediction and hidden state prediction is the mean squared error (MSE). The model is configured with 1,600 hidden neurons, a learning rate of 0.0001, and $N_1=100$ iteration steps during the writing phase. When reading the stored memory, $N_2=500$ iteration steps are utilized.

After completing the read operation, the saved tensor is reconstructed into an audio file, referred to as the recalled audio. To evaluate the similarity between the recalled audio and the original, we generate natural language captions and transcriptions of the speech content, respectively. For this purpose, we employ two pre-trained models: CLAP~\cite{elizalde2023clap}, which demonstrates state-of-the-art performance on the \texttt{ESC-50} dataset for environmental sound captioning, and Whisper~\cite{radford2023robust}, a robust speech recognition model. These models allow us to quantitatively and qualitatively assess the fidelity of the recalled audio in both environmental and speech-based tasks.

We evaluate the similarity between the original and recalled audio by analyzing both the semantic similarity of generated text pairs and the acoustic similarity of the audio signals. Captions for environmental sound pairs and recognized speech content for speech sound pairs are generated using the pre-trained models. To assess semantic similarity, we compute the cosine similarity of sentence embeddings, which range from 0 to 1, where higher values indicate greater similarity. The embeddings are generated using a MiniLM-based model~\cite{reimers-2019-sentence-bert}, a 6-layer miniature transformer optimized for efficiency and semantic representation.

% Given that deep learning models can be highly sensitive to small perturbations in input data, we also compute the Mel-Frequency Cepstral Coefficients (MFCC) distance $d_{\text{MFCC}}$ between audio pairs to provide an additional metric for evaluating audio similarity, which is calculated by Eq~(\ref{eq:MFCC_distance}).
% \begin{equation}
% d_{\text{MFCC}} = \frac{1}{T} \sum_{t=1}^{T} \left\| \mathbf{M}_{\text{recall}}(:, t) - \mathbf{M}_{\text{original}}(:, t) \right\|_2,
% \label{eq:MFCC_distance}
% \end{equation}
% where $\mathbf{M}_{\text{recall}}$ is the MFCC matrix of the recalled audio signal with dimensions $n \times T$ (where $n$ is the number of MFCC coefficients and $T$ is the number of time frames), and $\mathbf{M}_{\text{original}}$ is the MFCC matrix of the original audio signal with the same dimensions. $\left\| \cdot \right\|_2$ denotes the Euclidean norm (L2 norm) computed for each time frame $t$, $T$ is the total number of time frames, and $d_{\text{MFCC}}$ represents the average Euclidean distance, quantifying the overall difference between the two signals in the MFCC feature space.

\section{Results}
In this section, we first demonstrate the necessity of incorporating the close-loop form into the proposed framework. Subsequently, we analyze the recalled audio from both datasets using the previously introduced evaluation metrics.

\subsection{Functionality of the close-loop Model}

As speech data typically contains detailed and fine-grained audio content, we present two waveform examples in Fig.~\ref{fig:close_loop_figures} for analysis. For the example in Fig.~\ref{fig:close_loop_figures}(a), the Whisper model accurately recognizes the original audio as `I had wanted to get some picture books for Yulka and Antonia.' Similarly, in Fig.~\ref{fig:close_loop_figures}(b), the recognition result for the original audio is `I'm not a talker, you know, and, as the laws of gravitation forbid my soaring aloft anywhere', which is also accurate.

The upper panel of Fig.~\ref{fig:close_loop_figures}(a, b) compares the waveform of the original audio with the recalled audio generated without using the close-loop waveform. In contrast, the lower panel compares the original audio's waveform with the recalled audio produced using the close-loop waveform.

Fig.~\ref{fig:close_loop_figures}(a) clearly demonstrates that the close-loop approach significantly improves recall accuracy, as indicated by the closer alignment of the blue waveform with the red waveform. In the upper panel, the recalled audio's waveform exhibits repetitive patterns after the initial 0.2 seconds, containing no meaningful information and consisting predominantly of noise. Conversely, in the lower panel, the recalled audio's waveform more closely resembles the original, with similar slopes but an increased amplitude. This increase in amplitude likely results from additional echo and background noise introduced during the memory reading process. Despite these distortions, the Whisper recognition result remains consistent with the original audio transcription.

Fig.~\ref{fig:close_loop_figures}(b) further illustrates that the close-loop approach significantly enhances recall accuracy, as evidenced by the closer alignment of the blue waveform with the red waveform. In the upper panel, the recalled audio's waveform exhibits very low amplitude after the initial 0.2 seconds, containing predominantly noise and no meaningful information. In contrast, the lower panel shows that the recalled audio's waveform more closely matches the original, with similar slopes but reduced amplitude and a slight temporal shift of several milliseconds. Despite these distortions, the Whisper model's recognition result remains consistent with the original audio transcription.

To demonstrate the effectiveness of the proposed model in environmental sound classification, Fig.~\ref{fig:close_loop_figures_esc} presents the classification probabilities for all classes in the \texttt{ESC-50} dataset, comparing both the original sounds and the recalled sounds. Fig.~\ref{fig:close_loop_figures_esc}(a) illustrates that the CLAP model performs well on the dataset, as indicated by the diagonal elements in the heatmap being close to 1. However, certain classes such as `crickets', `helicopter', `hen', and `washing machine' exhibit relatively lower accuracy. Notably, the misclassified samples often shift to semantically similar classes—for instance, some `helicopter' sounds are classified as `airplane', which reflects the inherent ambiguity rather than pure misclassification.

For the recalled sounds generated by the model, most diagonal elements retain probabilities above 0.7, demonstrating the model's robust performance. Nonetheless, for sparsely distributed probability classes in the original dataset, there is an increase in misclassification across nearly all classes in the recalled sound, leading to a sparser overall distribution. Importantly, the close-loop feedback model significantly outperforms its counterpart without feedback, showcasing the benefits of the proposed approach.

\subsection{Evaluation of Recalled Environmental Sounds}
% \subsubsection{Semantic Similarity}

% \begin{figure*}[!t]
%     \centering
%     % Left figure
%     \includegraphics[width=0.48\textwidth, trim={0.4 0 0 28}, clip]{figures/Semantic Similarity.pdf}
%     % Right figure
%     \includegraphics[width=0.48\textwidth, trim={0.4 0 0 28}, clip]{figures/MFCC Distance.pdf}
%     \caption{(a) Boxplot of semantic similarities (SS) between original and recalled audio pairs in the \texttt{ESC-50} dataset. The x-axis represents the 50 class names, while the y-axis shows the semantic similarity scores. Black dots indicate the distribution of similarity scores for each class. (b) Boxplot of Mel-Frequency Cepstral Coefficient (MFCC) distances between original and recalled audio pairs in the \texttt{ESC-50} dataset. The x-axis represents the 50 class names, while the y-axis shows the MFCC distance values. Black dots indicate outliers in the distribution for each class.}
%     \label{fig:ESC_results}
% \end{figure*}

% \begin{figure}
%     \centering
%     \includegraphics[width=\linewidth, trim={0.4 0 0 28}, clip]{figures/Semantic Similarity.pdf}
%     \caption{Boxplot of semantic similarities (SS) between original and recalled audio pairs in the \texttt{ESC-50} dataset. The x-axis represents the 50 class names, while the y-axis shows the semantic similarity scores. Black dots indicate the distribution of similarity scores for each class.}
%     \label{fig:Semantic Similarity}
% \end{figure}

Fig.~\ref{fig:ESC_results} provides a comprehensive view of the class-wise semantic similarity distributions between original and recalled audio pairs in the \texttt{ESC-50} dataset. Each boxplot represents the variability of semantic similarity scores for a particular class, offering insights into how well the recalled captions align with their original counterparts. The majority of the classes achieve median semantic similarity scores above 0.7, indicating that the model generally performs well in maintaining semantic integrity during recall. 

The distributions reveal a notable variation in semantic recall consistency across different audio classes. Certain classes, such as `church bells', `clapping', `chirping birds', `siren', and `toilet flush', show tightly clustered distributions, reflected by narrow variance range. And the mean semantic similarities are all above 0.7. This indicates a high level of consistency in the semantic alignment for these sounds, suggesting that their auditory features are easier for the model to interpret and recall accurately. Such performance highlights the robustness of the model for these classes, potentially due to their distinct and repetitive auditory patterns.

Conversely, other classes, including `vacuum cleaner', `door wood creaks', `fireworks', `sneezing', `rooster', and `keyboard typing', exhibit significantly broader distributions. This variability points to challenges in achieving consistent semantic recall, which might arise from the inherent complexity or overlap of auditory characteristics within these categories. The variability could also indicate that the model struggles with sounds that lack clear defining features or have substantial intra-class diversity.

In addition to the variability within the distributions, outliers are present for several classes, visible as black dots beyond the whiskers of the boxplots. These outliers represent cases where the semantic similarity deviates significantly from the central tendency of the distribution, potentially signaling instances where the model failed to preserve the semantic content effectively. The presence of these outliers suggests areas where model improvements, such as enhanced feature extraction or better representation learning, may be needed to reduce inconsistencies.

\subsection{Evaluation of Recalled Speech Audio}

Fig.~\ref{fig:speech_results}(a) displays a box plot of semantic similarity (SS) scores, showing a wide range of values across the dataset. Most samples cluster around 1, with median above 0.8, indicating good alignment between the original speech and the model’s recall. However, a few outliers with lower scores reveal that some samples deviate significantly, suggesting challenges in retaining meaning in certain cases.

% Fig.~\ref{fig:speech_results}(b) presents a box plot for the MFCC distances, which reflect the acoustic fidelity of the model’s recall of speech. The majority of samples exhibit low distances, indicating effective retention of acoustic features. Nevertheless, the variability in distances suggests the model’s performance is not consistent for all samples.

To further analyze the model’s performance, two contrasting examples are highlighted. In the first case, the sample achieves a high semantic similarity (SS = 1.00) in Fig.~\ref{fig:speech_results}(b), showing that the model successfully retained both meaning and acoustic characteristics. In the second case, the semantic similarity is negative (SS = 0.39) in Fig.~\ref{fig:speech_results}(c). This indicates that, despite some acoustic fidelity, the model failed to retain the intended meaning.

These examples highlight the model’s strengths and weaknesses in handling semantic and acoustic information. While it performs well for many samples, the variability in results underscores the need for improvements to achieve consistent performance across both dimensions.

\section{Conclusion}
We proposed a general neural network framework based on closed-loop predictive coding for auditory working memory. The model effectively demonstrates sequential memory capabilities by predicting subsequent auditory segments, capturing both hidden states and output observations. Memory is encoded into the weight matrices, enabling stored information to be recalled with frozen weights. To evaluate the framework, we conducted experiments on two acoustic datasets, \texttt{ESC-50} and \texttt{LibriSpeech}. Notably, the classification accuracy remains largely preserved during recall. Furthermore, we introduced a novel metric—semantic similarity—to quantify the accuracy of the recalled audio. Experimental results indicate that the semantic similarity scores consistently exceed 0.7 across both datasets, underscoring the framework’s ability to retain meaningful auditory representations.

% \section*{Acknowledgment}

% The preferred spelling of the word ``acknowledgment'' in America is without 
% an ``e'' after the ``g''. Avoid the stilted expression ``one of us (R. B. 
% G.) thanks $\ldots$''. Instead, try ``R. B. G. thanks$\ldots$''. Put sponsor 
% acknowledgments in the unnumbered footnote on the first page.

\bibliographystyle{IEEEtran}
\bibliography{ref}

% \vspace{12pt}
% \color{red}

\end{document}